# State of health estimation for lithium-ion battery by combing incremental capacity analysis with Gaussian process regression

Xiaoyu Li[1,2,], Zhenpo Wang[1,2,*]

[1]National Engineering Laboratory for Electric Vehicles, School of Mechanical Engineering, Beijing Institute of Technology, Beijing 100081, China

[2]Collaborative Innovation Center of Electric Vehicles in Beijing, Beijing Institute of Technology, Beijing 100081, China

**Abstract:** The state of health for lithium battery is necessary to ensure the reliability and safety for battery energy storage system. Accurate prediction battery state of health plays an extremely important role in guaranteeing safety and minimizing the maintenance costs. However, the complex physicochemical characteristics of battery degradation cannot be obtained directly. Here a novel Gaussian process regression model based on partial incremental capacity curve is proposed. First, an advanced Gaussian filter method is applied to obtain the smoothing incremental capacity curves. The health indexes are then extracted from the partial incremental capacity curves as the input features of the proposed model. Otherwise, the mean and the covariance function of the proposed method are applied to predict battery state of health and the model uncertainty, respectively. Four aging datasets from NASA data repository are employed for demonstrating the predictive capability and efficacy of the degradation model using the proposed method. Besides, different initial health conditions of the tested batteries are used to verify the robustness and reliability of the proposed method. Results show that the proposed method can provide accurate and robust state of health estimation.

I. INTRODUCTION

Lithium-ion batteries have been regarded as the leading energy storage source for many electrification fields such as electric vehicles, micro-grids, and other consumer electronics, thanks to their excellent properties in self-discharge rate, lifespan, energy density, and power capability [1, 2]. However, the battery degradation with operation process would lead to capacity loss and accident-proneness. Therefore, knowledge of the present battery health status is essential to avoid battery running risk and ensure the battery reliability and safety[3, 4]. Among these problems, accurately online monitoring battery external characteristics and evaluating state of health (SOH) are huge challenging issues of the battery management system (BMS). The precise results of battery SOH estimation not only indicate battery aging level but also provide valuable guidance for reasonable using batteries [5, 6].

The SOH estimation has various definitions generally considering two aspects: capacity and internal resistance. The two definitions both reflect a proportion of the residue and rated value. Meanwhile, the battery end of life (EOL) is often defined as the available capacity takes up 80% rated capacity or the internal resistance increases to twice the rated internal resistance [7, 8]. Due to the complex and nonlinear electrochemical mechanisms and various application scenarios, it becomes very difficult to estimate the battery SOH accurately through ordinary measurement [9, 10]. Hence, there have been several methods put forward for battery SOH estimation, which can be generally grouped into three categories: empirical or semi-empirical models, electrochemical/physical-based, and data-driven approaches.

Empirical or semi-empirical models use laboratory test data-fitted equivalent circuit models (ECMs) or simple battery degradation models to depict the battery characteristics. Advanced observers are synthesized to capture the bulk capacity and/or internal resistance that are indicative of battery health. In [11,12], genetic algorithm (GA) and recursive least square (RLS) are applied to obtain the internal resistance based on standard equivalent circuit model (ECM). Additionally, the pulse approaches are employed to track the internal resistance under different depth of discharging (DOD) [13]. Then the relationship between the real-time resistance and two fixed resistances (the maximum and minimum resistances) indicates the battery degradation condition. Based on experimental battery capacity data, some fitting methods and prediction algorithms are harnessed to evaluate the actual capacity, such as an empirical exponential method [14], multivariate adaptive regression splines [15]. However, their accuracy is limited because the battery degradation is generally affected by some unknown stressing factors. Hence, the accuracy of battery SOH estimation highly depends on the adaptability and robustness of the applied models.

Electrochemical/physical-based methods generally leverage the mathematical and physical techniques to simplify electrochemical models to delineate battery health levels over the battery lifetime. These models can give insights into the electrochemical process inside the battery and the key chemistry and physics mechanisms such as diffusion, migration, and reaction kinetics [16, 17]. The growth of solid electrolyte interface (SEI) film is harmful side-reaction and widely regarded as the primary cause for Li inventory and active material loss. These respectively contribute to capacity fade in storage and cycling conditions [18]. However, the electrochemical models are usually coupled partial differential equations (PDEs) that lead to the computational intensity required for PDE-based model calculation hinders its feasibility.

Data-driven methods get more attention due to their model-free characteristics [19]. These methods build battery degradation through mapping external characteristics to battery capacity loss. Otherwise, some methods focus on the tendency of global degradation and the previous capacities as input components. This results in a variety of approaches using support vector machines (SVMs) [20, 21], Bayesian networks [22, 23], Autoregressive models [24], and Gaussian process regression (GPR) [25, 26]. Although the data-driven methods have good nonlinear property and good estimation accuracy, the methods also require high-quality datasets for training purpose.

At present, some research and literature find that battery degradation is closely related to the terminal voltage during the charging process. In [27], the charging terminal voltage is regarded as features to reflect battery remaining capacity. In-depth analysis, during the charging process, combining capacity and voltage to reflect battery degradation is rewarding things. Therefore, the incremental capacity analysis (ICA) is proposed as the features to evaluate battery health condition. Incremental capacity (IC) can be calculated by differential the charging/discharging capacity over the voltage evolution. The IC curve has high resolution for the charging/discharging voltage plateau region, and what's more, aging mechanisms can be extracted from the peak amplitude and position of the curve. Otherwise, the ICA studies have been validated for on-line SOH estimation for various perspectives such as area, position, and gradient [5, 28]. In [29], based on IC curve, feasibility and accuracy of the degradation model is established using Gaussian function. The battery SOH estimation is achieved by analysis the degradation model of capacity. To sum up, the ICA has excellent performances in achieving SOH estimation. However, there exists a difficult problem for capturing the peak in the IC curves, where the peak is submerged by measurement noises.

This paper analyzes the above mentioned technical difficulties for the inferior quality datasets of the data-driven method and the noise sensitive IC curve. To solve these problems, an advanced smoothing method based on Gaussian filter algorithm is proposed to obtain a smooth IC curve at first, and a partial region of IC curve is selected to extract battery degradation features using a linear interpolation method. Correspondingly, the features of the region are defined as health performance indicators (HPIs) that are regarded as high-quality datasets for battery degradation models. Secondly, the GPR is modeling for battery degradation based on Bayesian method. The important aspect of the GPR model is not only evaluating the battery capacity loss but also expressing the uncertainty of estimated results using the confidence interval with upper and lower bounds. Finally, the effectiveness and robustness of the proposed method is experimentally verified using four aging datasets are extracted from NASA data repository.

The structure of this paper is arranged as follows: Section II introduces the ICA method and analysis of battery aging phenomena. Section III describes the proposed GPR method for battery SOH estimation. Section IV presents the estimation results and discussions. The key conclusions are summarized in Section V.

## II. BATTERY AGING PHENOMENA AND INCREMENTAL CAPACITY ANALYSIS

Four aging battery datasets from NASA data repository are considered in this section. First, the battery degradation phenomena are analyzed considering the changes of voltage and capacity in the section A. After that, the IC curves are extracted from the voltage evolution process during the constant-current charging step and the relational features of IC curves with battery SOH are obtained in section B.

*A. Battery aging phenomena analysis*

Experimental datasets are obtained from the NASA Ames Prognostics Center of Excellence and four batteries (labeled as No.5, 6, 7, and 18) are the second-generation, Gen 2, 18650-size LIBs produced by Idaho National Laboratory [30]. All the batteries are carried out recurrently through three different operational profiles at room temperature 24°C. Specifically, the three operational models consist of constant-current and constant-voltage charge (CC-CV) mode, constant current (CC) discharge and impedance measurement modes. The batteries cycle and capacity degradation profiles are shown in Fig. 1. From the Fig. 1(a), the charging process contains the CC and CV mode, where the constant current and the charging cut-off voltage are set as 1.5A and 4.2V, respectively. Once the charging voltage reaches to cut-off voltage, the charging process turns into CV mode until the charging current drops to 20mA. Then constant current discharging is continued with 2A. The four batteries have the same charging process but they have different discharging cut-off voltages those are 2.7, 2.5, and 2.2V for corresponding batteries. The detailed experimental conditions are listed in Table. I. The degradation tendencies of battery capacities under different cycle numbers are described in Fig. 1(b). The battery degradation capacities are in unsteady decreasing trends with cycle numbers and they have some regeneration capacities during the discharging process. This phenomenon is gaining the difficult to evaluate the battery SOH because the battery capacities have nonlinear relationship with the battery cycle number. Hence, some other great correlative parameters are considered to improve the accuracy of battery SOH estimation. The IC curves are applied for battery SOH estimation and the IC curves are detailed analysis in the next section.

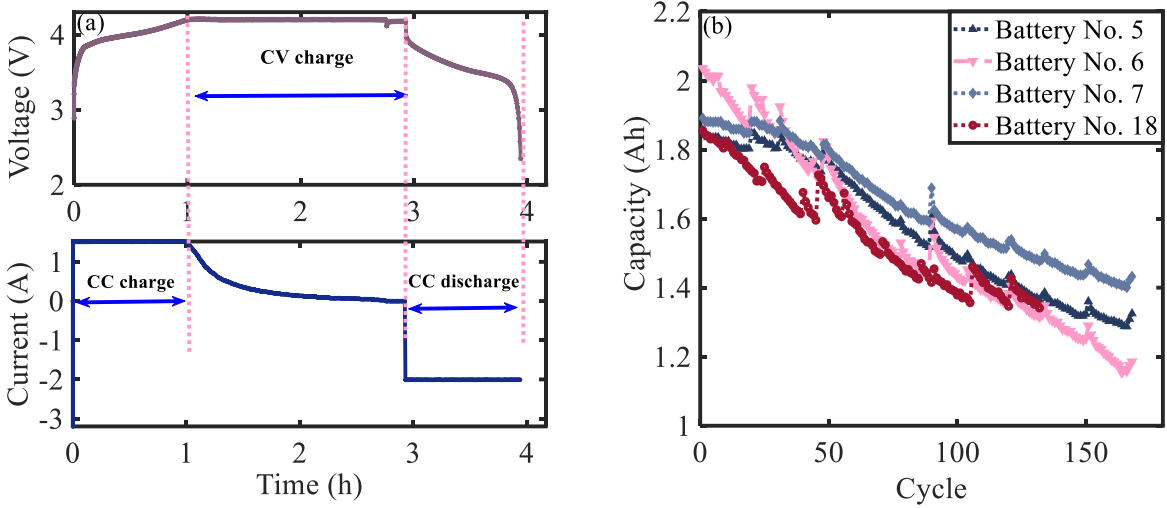

Fig. 1. The batteries cycle and capacity degradation profiles: (a) The voltage and current during a complete test cycle; (b) capacity degradation trends of the four batteries.

Table. I. The four batteries' specific cycle condition

| Battery label | Batteries' cycle condition | | | | |
|---|---|---|---|---|---|
| | Charging cut-off voltage (V) | Discharging cut-off voltage (V) | Charging constant current (A) | Discharging current (A) | Temperature (°C) |
| No. 5 | 4.2 | 2.7 | 1.5 | 2 | 24 |
| No. 6 | 4.2 | 2.5 | 1.5 | 2 | 24 |
| No. 7 | 4.2 | 2.2 | 1.5 | 2 | 24 |
| No. 18 | 4.2 | 2.5 | 1.5 | 2 | 24 |

B. *Incremental capacity curve analysis*

The IC curve is widely regarded as an effective tool to analyze the capacity loss of battery and obtained from charging process under constant-current regime by using differential equation. Specifically, the IC curve is described as the amount of incremental capacity over a successive voltage step. Thus, the battery charging capacity and voltage should be known before calculation IC values. Owing to constant current, the capacity and voltage can be calculated as:

$$Q = It \tag{1}$$

$$V = f(Q), Q = f^{-1}(V) \tag{2}$$

where *t* is charging time and *I* refers to charging current. Based on the Eq. (1), the IC curve can be expressed as follows,

$$(f^{-1})' = \frac{dQ}{dV} = \frac{I \cdot dt}{dV} = I \cdot \frac{dt}{dV} \tag{3}$$

Obviously, the IC values are inversely proportional to $dV/dt$. The voltage and current data are usually collected from the commercial BMS. Hence the form of incremental capacity is transformed in voltage step forward and the range covers the two cut-off voltages.

In practical, it is difficult to obtain the peaks based on incremental capacity method, because the parameter values are vulnerable to perturbation of measurement noise during the charging plateau region, as shown in Fig. 1(a). To address this problem, some effective filter algorithms and machine learning are proposed to deal with those problems such as support vector machine [31] and Gaussian function and Lorentzian function [32]. Those filter methods have better accuracy for their research objects. Here, the moving average (MA) and Gaussian filter (GS) methods are chosen to smooth the intact IC curve and the IC curve is extracted from the first cycle of the labeled No. 6 battery.

MA method is based on the statistics regularity and widely applied for signal process field. The method selects the continued historical data as a series of numbers and a fixed subset size with *N* sample points. The first filtered point of the method is the average of the initial fixed subset with *N* sample points. After the sample point update, the first sample point of the previous series is removed and the new sample point is added the residual *N*-1 sample points. Then the filter point is updated using the average of the new series. The MA method is expressed as follows,

$$y(i) = \frac{1}{N} \sum_{j=0}^{N-1} x(i+j) \tag{4}$$

where $x(\cdot)$ is the sample point as the input signal and $y(\cdot)$ is the output signal. *N* is the fixed size of the series. With the filter method, the random measured noise can be eliminated when the impulse response is sampled.

However, the filter results of the MA method is closely depend on fixed size. Roughly, the bigger fixed size is, the smoother IC curves we can obtain. But the bigger fixed size will deform the IC curve and can therefore lead to false interpretation of the important features. Nevertheless, the size should be controlled within a suitable range and it is time-consuming to choose the size.

GS method is regarded as the ideal time domain filter for effective separation low-frequency signals from the higher frequency noises. The method has wide application in image and signal process for one- dimension or high-dimension signals. The GS method has the characteristic of a Gaussian distribution, which is expressed as

$$G(x) = \frac{1}{\sigma\sqrt{2\pi}} exp\left(\frac{-(x-\mu)^2}{2\sigma^2}\right) \qquad (5)$$

where $\mu$ and $\sigma$ are the mean value and standard deviation, respectively. When the GS method is employed to smooth the IC curve, each sample point will be filtered by a weighted average of its neighbors. Therefore, the nearest points have more influence on the results and the distant points have less importance. Generally, when the GS is applied for smoothing, $\mu$ is set to smaller number because an amount of $\mu/2$ sample points cannot filter for the head and the tail of the sample series. Here $\sigma$ can serve as a parameter to control how much smooth for the final curve, more specifically, how big the fixed size for averaging. Too small fixed size cannot meet the desired smoothing effect, while too large size may lead to losing important information. The parameters of the GS method are selected through comparing with the results of the MA filter and the $\mu$ and $\sigma$ are set as 17 and 5, respectively, in this study. As shown in Fig. 2(b), the two methods mentioned above are compared for filtering the initial IC curve. From the Fig. 2(b), the GS method has excellent capability for smoothing the IC curve. Compared with the MA method, the GS method can identify the peak points of the IC curve easily and clearly. Hence, the GS method is employed to obtain the features of battery degradation through the IC curve in this study.

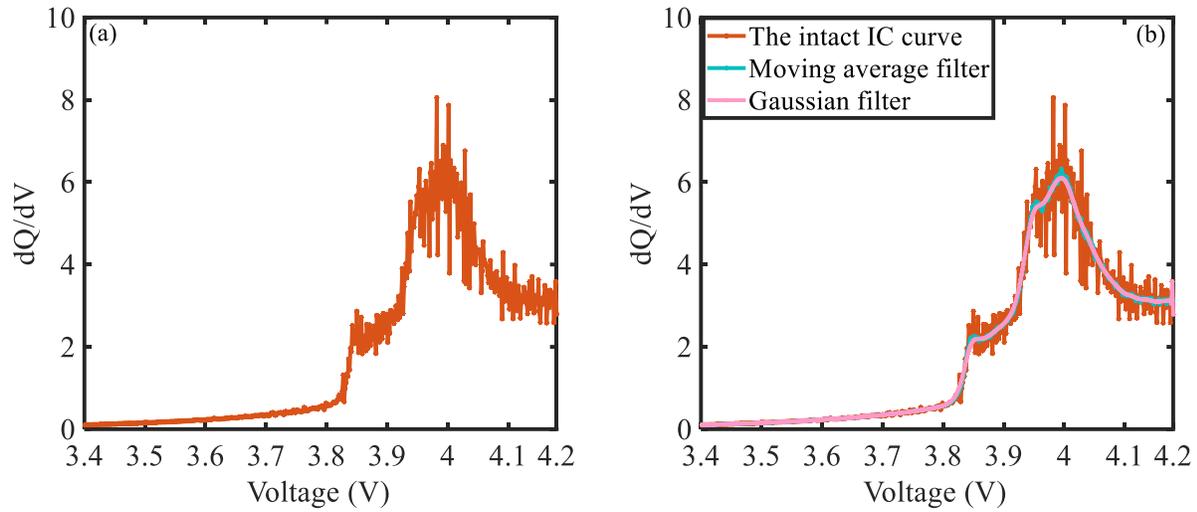

Fig. 2. The IC curve of battery No. 6. (a) The initial IC curve with noise; (b) the comparison of IC curves smoothed by Gaussian filter and Moving average method

*C. Features extraction of battery degradation*

For studying battery health levels, a number of charging/discharging cycles are carried out based on the test schedule mentioned above. With increasing battery cycles, the anode/cathode's materials and the electrolyte solvent of battery gradually change because of complex physical and chemical characteristics which, in turn, leads to an increment of the internal resistance and a decrement of the available capacity. In practical operation for electrical applications and EVs, the available battery capacity reaches to the 80% of rated capacity is regarded as battery EOL. Typically, the EOL means a power of battery cannot meet the EVs' requirement and the battery need to be replaced.

Here, the experimental No. 6 battery data are employed to illustrate battery degradation. The battery degradation data are extracted from experimental data at intervals of 30 cycles, as shown in Fig. 3. It is worthwhile to note that the upper cut-off voltage is fixed at 4.2V during the charging process, but the degradation battery has less charging time that compared with the fresh battery. From the Fig. 3(a), the fresh battery complete the CC step that needs 3600s or so, while the aged battery just needs about 1800s and they both from the same charging point. Meanwhile, a large proportion of battery charged capacity of the different health levels is covered in this CC charging mode, as shown in Fig. 3(b). The terminal voltage changes from 3.8V to 4.2V and the range takes up 25% of the working voltage range (3.2V to 4.2V), however, the charged capacity accounts for approximately 60% of rated capacity. Hence, the IC curves can be regarded as an excellent analysis method for identifying different aging battery levels. The whole IC curves of the battery of different cycle numbers are plotted in Fig. 3(c). Generally, the IC curve contains significance features to research battery health levels through analysis the change trends of IC curves. From Fig. 3(c), the shapes of IC curves have no remarkable changes in a voltage range from 3.4V to 3.8V. While the IC curves have distinct changes between 3.8V and 4.1V and this region covers the dominating charged capacity for the battery of different health levels. Hence, the voltage range from 3.8V to 4.1V is regarded as important area for identifying the battery degradation. The battery HPIs are extracted from this region at the interval 30mV, as shown in Fig. 3(d). Therefore, the 11 features are obtained and applied to evaluate the battery health levels.

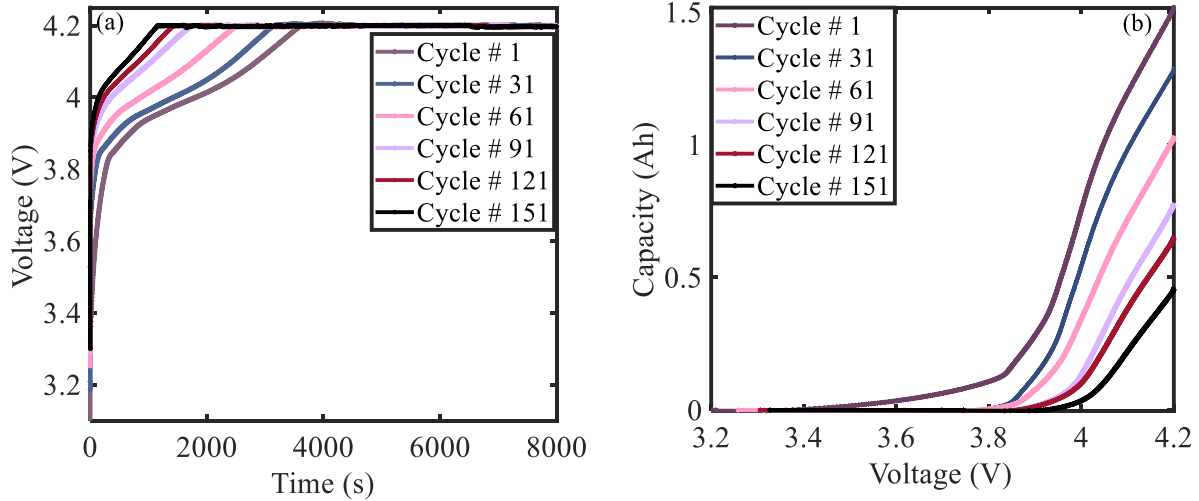

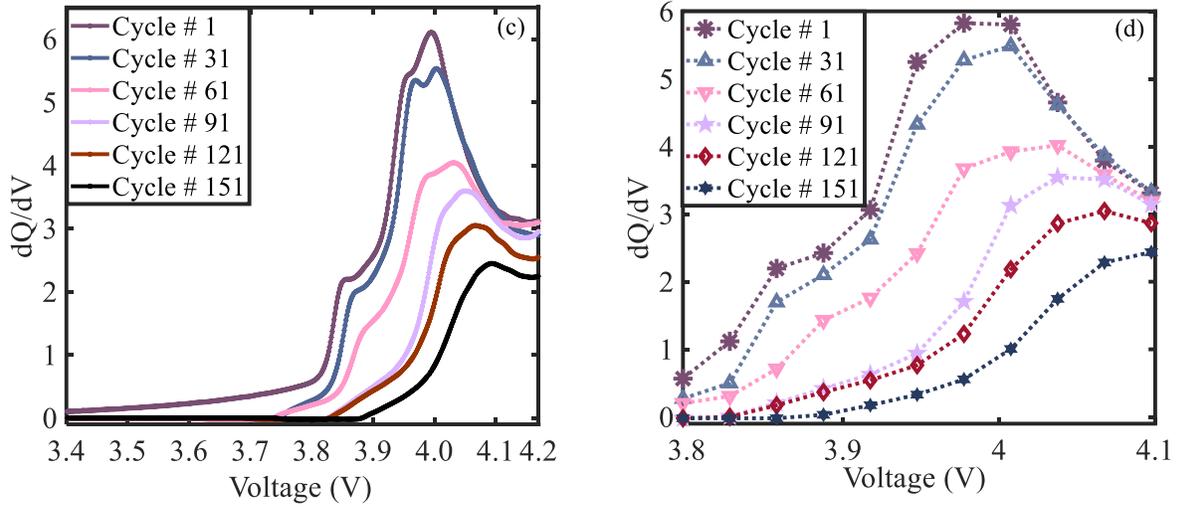

Fig. 3. The battery parameters' evolutions of the No. 6 battery under different cycles. (a) The voltage evolutions during the charging process; (b) Actual charged capacities in CC mode; (c) The complete IC curves with different cycles; (d) The significance evaluation features for battery degradation.

## III. METHODOLOGY

The GPR algorithm has been widely applied in machine learning and statistics fields due to the advantage of convenient properties for building model without specific functional form [25, 33]. Rather than other regression methods need to detailed parameters for modeling, the GPR uses probabilistic method to train the sample data. GPR is determined by a mean function and a covariance function and the Bayesian inference is used to obtain the hypothesis of the posterior probability. In this study, the goal of the GPR algorithm is to model the mapping from input features to output battery SOH.

### A. Description of the GPR algorithm

The GPR is based on the concept of the Gaussian distribution and can be extended multivariate Gaussian distributions to infinite-dimensionality [33, 34]. Here, let $D = (X, y)$ denotes a training dataset, where $X$ is the input data and $y$ regards as the output variable. The input $X$ consists of $D$-dimensional $N$ input vectors $X = \{x_1^D, x_2^D, x_3^D, \cdots x_N^D,\}$ and its multivariable Gaussian distribution is defined as $f(x)$ with the mean function

$m(x)$ and covariance function $k_f(x, x')$. The properties of $f(x)$ can be fully described using the mean function $m(x)$ and covariance function $k_f(x, x')$ as follows,

$$\begin{cases} m(x) = E(f(x)) \\ k_f(x, x') = E[(m(x) - f(x))(m(x') - f(x'))] \end{cases} \quad (6)$$

The corresponding output $y$ vectors $y = \{y_1, y_2, y_3, \cdots y_N\}$. It is hypothesis that there exists an underlying latent function $f(\cdot)$, which maps the inputs, $X$, to the corresponding output values, $y$:

$$y = f(X) + \varepsilon \quad (7)$$

where $\varepsilon$ is zero-mean additive Gaussian noise with variance $\sigma^2$, $\varepsilon \sim N(0, \sigma^2)$. For different input series data, the noise vector $\varepsilon = \{\varepsilon_1, \varepsilon_2, \varepsilon_3, \cdots \varepsilon_N\}$ forms an independent and identically distributed series. It is worthwhile noting that the latent function $f(X)$ according to the multivariate Gaussian distribution. Hence the latent function is related to the covariance function based on the Eq. (6). For the GPR algorithm, the kernel function $k_f(x, x')$ plays an important role because it can obtain the prior assumptions for the properties of the underlying latent function. Here the kernel function is calculated using the squared exponential covariance (SE) function, as described in Eq. (8).

$$k_f(x, x') = \sigma_f^2 \exp\left(\frac{-(x - x')^2}{2l^2}\right) \quad (8)$$

Considering each input vector contains $D$-dimensional feature data, the Eq. (8) can be rewritten as follows,

$$k_f(x_i, x_j) = \sigma_f^2 \exp\left(-\frac{1}{2} \sum_{d=1}^{D} \frac{(x_i^d - x_j^d)^2}{l_d^2}\right) \quad (9)$$

where $x_i^d$ and $x_j^d$ represent the $d$-th features of the input vectors $x_i$ and $x_j$, respectively. The parameters $\sigma_f$ and $l_d$ form the hyper-parameters matrix $\theta = [\sigma_f, l_1, l_2 \cdots l_d]^T$. Specifically, $\sigma_f^2$ controls the variation of the underlying latent function and $l_d$ denotes the characteristic length scale of each input vector. The parameter $l_d$ can determinate the importance between the input variable and the output targets. Normally, the smaller the parameter value indicates that the more important input vector with the corresponding output. Here, the noise covariance matrix $\sigma_n^2 I$ is added into the Eq. (9) as below,

$$k(x_i, x_j) = k_f(x_i, x_j) + \sigma_n^2 I \tag{10}$$

Considering the input vector $X$ and the latent function $f$, the distribution of $y$ is expressed as follows,

$$p(y|f, X) = N(f, \sigma_n^2 I) \tag{11}$$

where $I$ is a n-dimensional unit matrix. The marginal distribution of $y$ can be given by

$$p(y|X) = \int p(y|f, X) p(f|X) df = N(0, K + \sigma_n^2 I) \tag{12}$$

where the $K$ is kernel matrix and the distribution of $p(f|X)$ can be explained as $N(0, K)$. Then, the marginal log-likelihood of $y$ can be expressed as,

$$\log p(y|X, \Theta) = -\frac{1}{2} y^T (K + \sigma_n^2 I)^{-1} y - \frac{1}{2} |K + \sigma_n^2 I| - \frac{n}{2} \log 2\pi \tag{13}$$

Here the gradient method is applied to optimize the hyper-parameters. The basic thought of gradient method is solving the maximum value of objective function through taking the derivative of the marginal log-likelihood function. Hence, the calculation process of the gradient method for the marginal log-likelihood function in the following,

$$\begin{cases} \dfrac{\partial \log p(y|X, \Theta)}{\partial \theta_i} = -\dfrac{1}{2} \text{tr}(\phi^{-1} \dfrac{\partial \phi}{\partial \theta_i}) + \dfrac{1}{2} y^T \phi^{-1} \dfrac{\partial \phi}{\partial \theta_i} \phi^{-1} y \\ \phi = K + \sigma_n^2 I \end{cases} \tag{14}$$

where $\theta_i$ is an element of the hyper-parameter $\Theta$. It is worthwhile noting that in the computation process of the Eq. (14) involves an inversion for the matrix $\phi$ that a time-consuming process. Therefore, it needs to consider a simple implementation of the GPR for hundreds and thousands of training datasets. Here the sparse approximations propose to handle the GPR by using a small representative subset for training datasets.

Otherwise, the objective function is a nonconvex function for solving the hyper-parameters in general, and hence, the gradient method may converge to the local optimum. For the sake of this problem, the different initial points are applied for gradient-based optimization and the largest results of marginal log-likelihood are chosen. When determining the final the hyper-parameters, the model can be used to predict the test datasets. The joint prior distribution of the training output $y$ and the test output $y^*$ is deduced and expressed as

$$p(y, y^* | X, x^*, \Theta) = N\left(\begin{bmatrix} 0 \\ 0 \end{bmatrix}, \begin{bmatrix} \phi & k^* \\ (k^*)^T & k^{**} + \sigma_n^2 \end{bmatrix}\right) \tag{15}$$

where $x^*$ is the new datasets for input training. The kernel functions of $k^*$ and $k^{**}$ are represented as $k^* = [k(x_1, x^*), \cdots, (x_n, x^*)]^T$ and $k^{**} = k_f(x^*, x^*)$, respectively. The main task of the GPR is to predict the posteriori distribution of the test output datasets $y^*$ by computing the $p(y^*|x,y,x^*)$. Based on the Eq. (15), the posteriori distribution can be derived as

$$p(y^* | x, y, x^*) = N(y^* | \bar{y}^*, \text{cov}(y^*)) \tag{16}$$

where the mean $\bar{y}^*$ and covariance $\text{cov}(y^*)$ of the predictive distribution are listed as follows,

$$\bar{y}^* = (k^*)^T \phi^{-1} y \tag{17}$$

$$\text{cov}(y^*) = \sigma_n^2 + k^{**} - (k^*)^T \phi^{-1} k^* \tag{18}$$

At this point, a complete GPR modeling process is finished. The output mean is regarded as effective estimation result of the test datasets. In addition, the variance of the predictive distribution $\text{cov}(y^*)$ in (18) seen as a measure of the reliability for the test output. The GPR is used to build the battery SOH estimation model, as described next.

### B. GPR modeling for battery SOH estimation

The main goal of this study is to establish a reliable model of battery capacity degradation using the direct measured parameters such as voltage and current. Here the battery degradation model extracts the features from the partial IC curves that comprehensive the voltage and current parameters. The features are obtained from each charging process of continued cycle test. Therefore, the input datasets can be seen as a time series and the sample interval is per cycle.

According to the third part of section II, the input $X$ consists of $D$-dimensional $N$ input vectors can be specific defined as 11 dimensions and the $N$ is relative to the cycle number. The input datasets can be rewritten as $X = [(x_1^1, x_1^2, \cdots x_1^{11}), (x_2^1, x_2^2, \cdots x_2^{11}) \cdots, (x_N^1, x_N^2, \cdots x_N^{11})]^T$. The output datasets consist of the SOH values for the tested batteries.

In order to evaluate the performance of battery model, the 95% confidence interval (CI) is computed by using Eqs. (17) and (18) as follows,

$$95\%CI = \bar{y}^* \pm 1.96 \times \text{cov}(y^*) \tag{19}$$

where the 95% CI represents the under-or over-confident in this range of the results of battery health estimation, respectively. Otherwise, the mean absolute error (MAE) and the root mean square error (RMSE) are proposed to take a quantitative analysis of the estimation results, define as

$$MAE = \frac{1}{N} \sum_{i=1}^{N} | y_i - \bar{y}_i^* | \tag{20}$$

$$RMSE = \sqrt{\frac{1}{N} \sum_{i=1}^{N} (y_i - \bar{y}_i^*)^2} \tag{21}$$

where $\bar{y}^*$ is the estimation result and the y is the real measured battery SOH value. It is worth noting that $N$ is the total number of battery cycle test. Because the MAE and RMSE are absolute values, they can be compared the performances of the proposed method with different datasets that extract from different batteries.

The complete structure of the battery SOH estimation based on GPR is shown in Fig. 4. First, the experimental voltage and current data are obtained from the NASA database. The features are extracted from the battery IC curves, which are filtered by using Gaussian filter method. Then the input datasets are used to build the GPR-based battery degradation model. At last, the test datasets are used to verify the reliability of the model and the error analysis method is used to quantify the accuracy of the model.

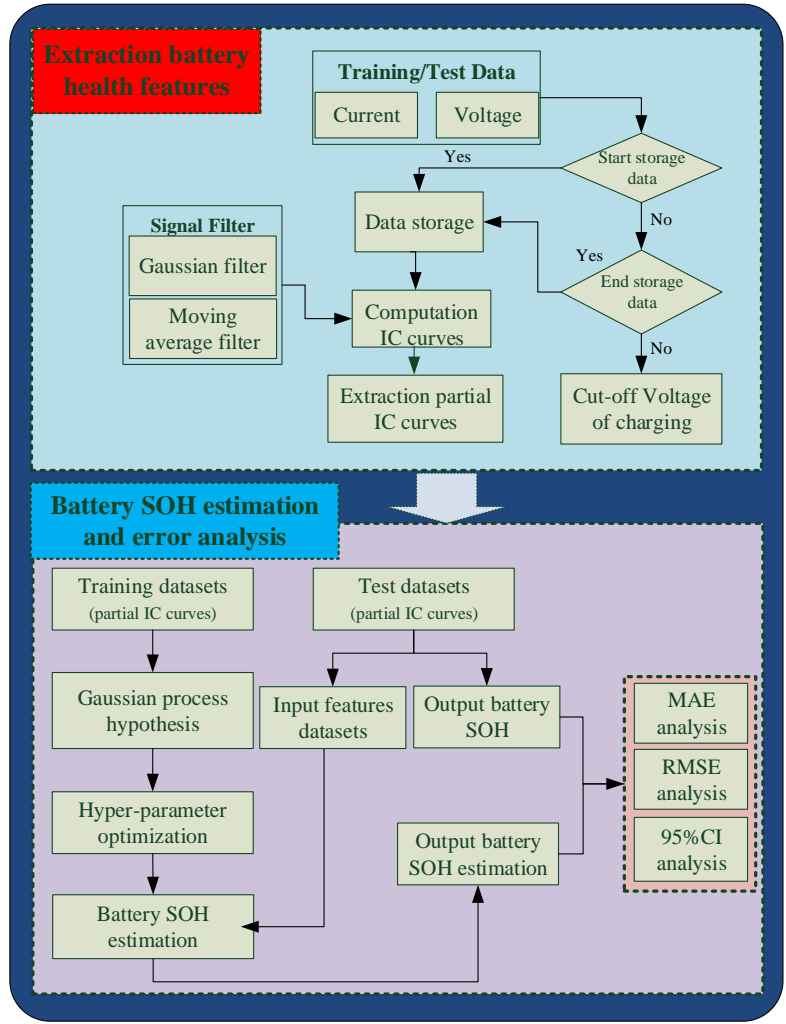

Fig. 4. The flowchart of the battery SOH estimation based GPR.

IV. BATTERY SOH ESTIMATION RESULTS AND DISCUSSION

In this section, the battery SOH estimation results of the GPR-based model are analyzed and verified. First, the four batteries labeled No.5, 6, 7 and 18 are used to verify the accuracy and effectiveness of the proposed battery degradation model. Then the different initial health conditions are set to further verify the reliability and robustness of the proposed model.

*A. Battery SOH estimation results based on GPR model*

Based on the proposed GPR battery degradation model, four batteries data of NASA database with different cycle test conditions are applied to verify the battery SOH estimation accuracy. During the modeling process, the first 55% tested data of each battery are regarded as the training datasets and the residual cycle data as testing datasets. The SOH estimation results and the relative errors for battery No. 5, 6, 7 and 18 are shown in Fig. 5. It is worth noting that, before the SOH estimation, the initial SOHs for the four batteries set as 1 according to the basic SOH definition. The legend of "Real SOH" is regarded as standard battery SOH according to test data, and the legend of "Estimation SOH" refers to the results of the SOH estimation with the proposed GPR-based battery degradation model. The "95% Confidence Interval" stands for the uncertainty of SOH estimation that can be divided into two parts such as the beyond and less than the estimation. Here in order to further verify the effectiveness of the battery SOH estimation model, the mean error and RMSE error analysis methods are used, as shown in Fig. 6.

As illustrated in Fig. 5(a-b), the results of battery SOH estimation and relative error analysis for the battery No. 5 that the discharging cut-off voltage is 2.7V for each cycle test. Under this cycling regime, the battery SOH estimation is in general similar with the actual application. During the model training process, the SOH estimation results have better accuracy, while the testing process the accuracy is decreasing and confidence intervals larger. However, the confidence intervals do accurately reflect the model uncertainty and the error-bars are used to analyze the model accuracy. From the Fig. 5(b), the errors are within 2% for the training process and that show surprisingly accurate SOH estimation almost within 4% during the test process. Fig. 5(c) shows the results of battery SOH estimation for the battery No. 6 that the discharging cut-off voltage is 2.5V. According to the 95% CI, it indicates that the battery degradation model has higher reliability both in training and testing periods. From the Fig. 5(d), the relative errors of the model are within 2% except two points because of the capacities impulsion regeneration. From the Figs. 5(e) and (f), the results and relative errors of the SOH estimation for battery No. 7 demonstrate that the model has high reliability and accuracy. The capacity degradation trend of battery No. 18 has large fluctuation around the EOL compared with other three batteries, as shown in Fig. 5(g). The confidence interval is in a relative narrow range that reflects the model has high reliability. Fig. 5(h) shows the relative errors are main distribution from -2% to 2% and the results also indicate that the model has high accuracy.

To analyze the performance of the battery degradation model, the MAE and RMSE are plotted in Fig. 6. As the name suggests, the MAE can reflect the average of the absolute difference between the real battery SOH and the battery estimation SOH. The MAE is not identical to RMSE, which is the square root of the average of squared errors and sensitive to outliers. Fig. 6(a) shows the maximum MAE is the battery No. 5 about 1.2% and the MAE of the residual batteries are around 0.4%. While the RMSE of the four batteries are all around 1% and the maximum and minimum RMSE are 1.38% and 0.78%, respectively, as shown in Fig. 6(b). The results from MAE and RMSE both indicate that the proposed method for battery SOH estimation has high accuracy and reliability.

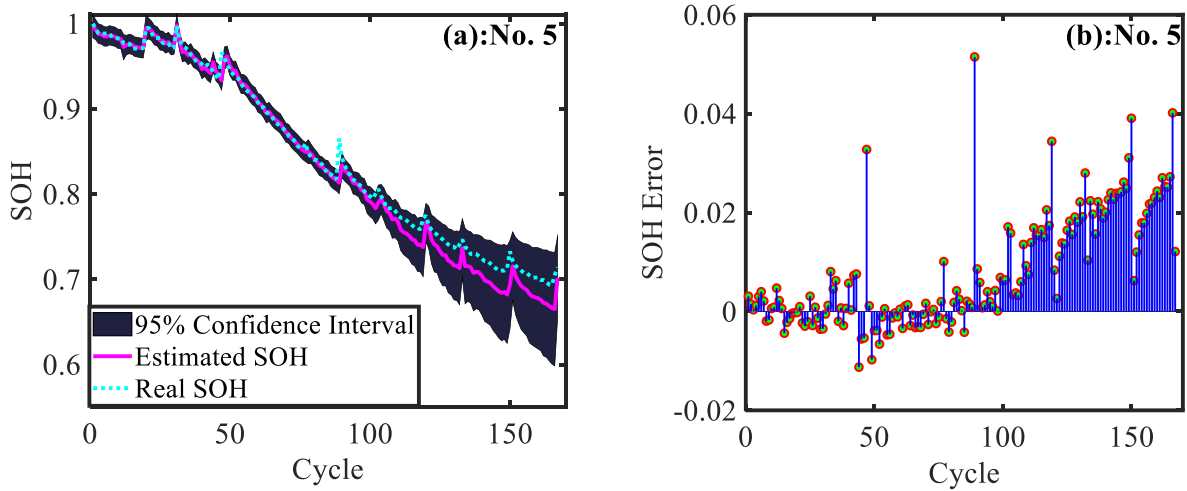

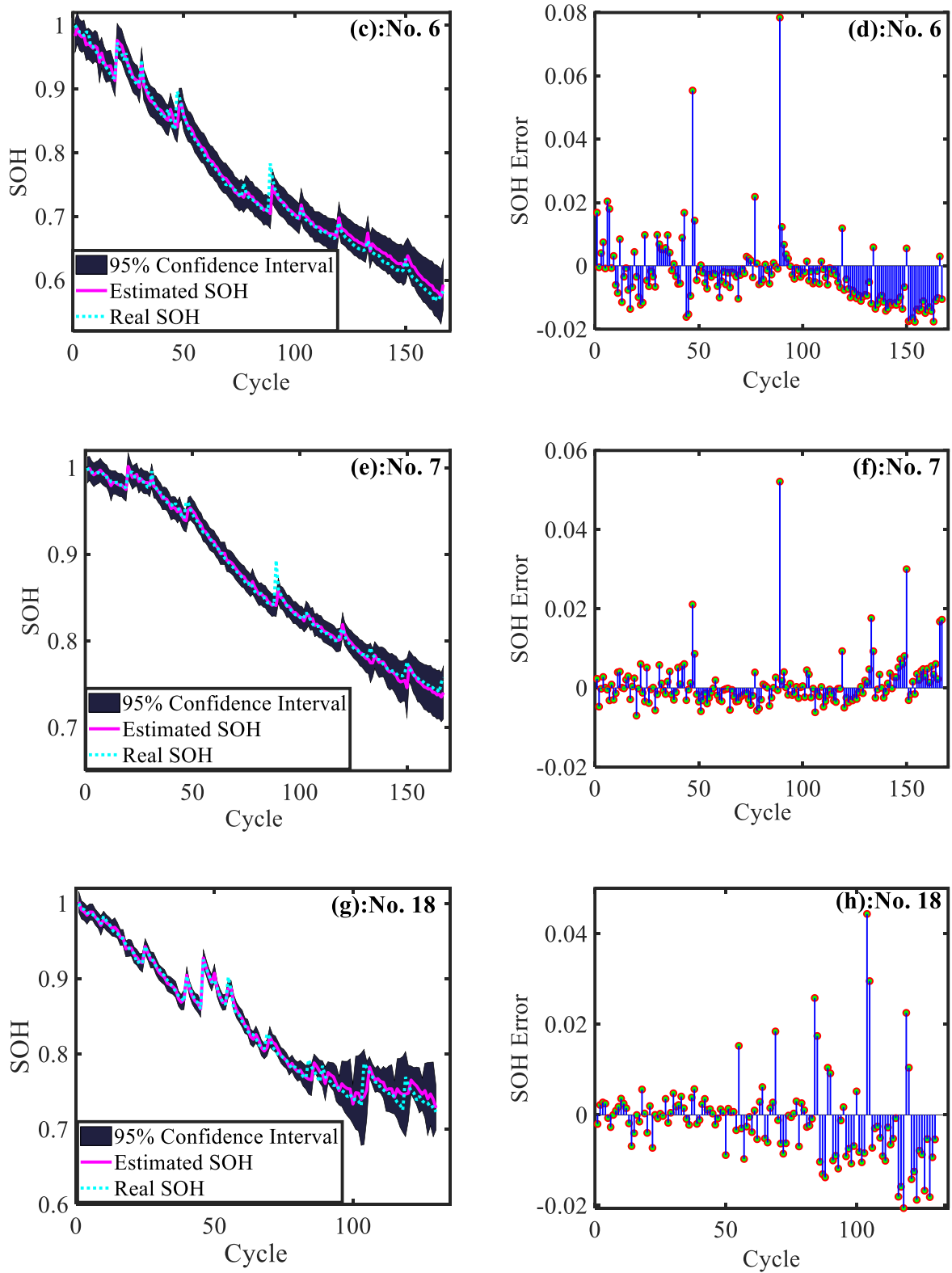

Fig. 5. Battery SOH estimation based on GPR model: (a) SOH estimation result for battery No. 5. (b) SOH estimation error for battery No. 5. (c) SOH estimation result for battery No. 6. (d) SOH estimation error for battery No. 6. (e) SOH

estimation result for battery No. 7. (f) SOH estimation error for battery No. 7. (g) SOH estimation result for battery No. 18. (h) SOH estimation error for battery No. 18.

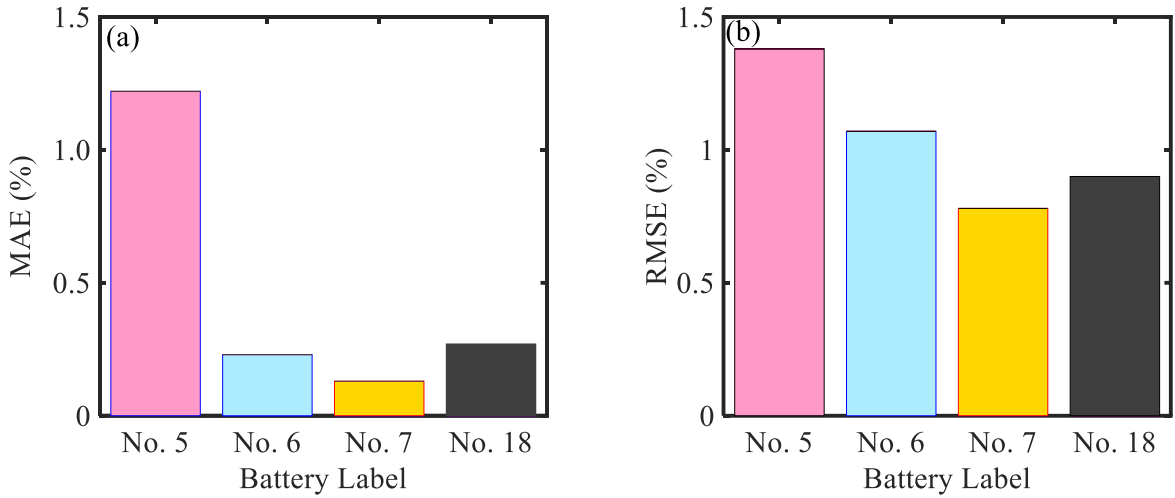

Fig. 6. Battery SOH estimation error analysis: (a) The MAE for the tested batteries. (b) The RMSE values for the tested batteries.

*B. Verification of GPR model robustness based on different initial conditions*

In this section, the robustness and reliability of the proposed battery degradation model have been verified by using the different initial health levels of the four tested batteries. Here all the datasets of tested batteries start from the 30th cycle and the first 60 cycles are selected as the training datasets. Fig. 7 shows the SOH estimation results and the relative errors for battery No. 5, 6, 7 and 18. The results of battery SOH estimation for the battery No. 5 is presented in Fig. 7(a). The SOH estimation results have better accuracy with smaller confidence intervals. From the Fig. 7(b), the error-bars are used to analyze the model accuracy that show surprisingly accurate SOH estimation almost within 1% during the whole process. Fig. 7(c) shows the results of battery SOH estimation for the battery No. 6 that in different health levels. According to the 95% CI, it indicates that the battery degradation model has higher reliability in training and testing periods. From the Fig. 7(d), the relative errors of the model are within -2% except two points are close to 8% because the capacities impulsion regeneration. From the Fig. 7(e-h), the results and relative errors of the SOH estimation for batteries No. 7 and No. 18 both demonstrate that the proposed method has high reliability and accuracy.

Fig. 8 shows the results of the error analysis of the four datasets. From Fig. 8(a), the maximum MAE is the battery No. 18 about 1.28% and the minimum MAE is around 0.25% for battery No. 5. The minimum RMSE of the battery No. 5 is 0.68% and the residual three values for corresponding batteries are all around 1.1%, as shown in Fig. 8(b). Summary, the proposed method can provide a reliable and robust SOH estimation according to the different initial health levels.

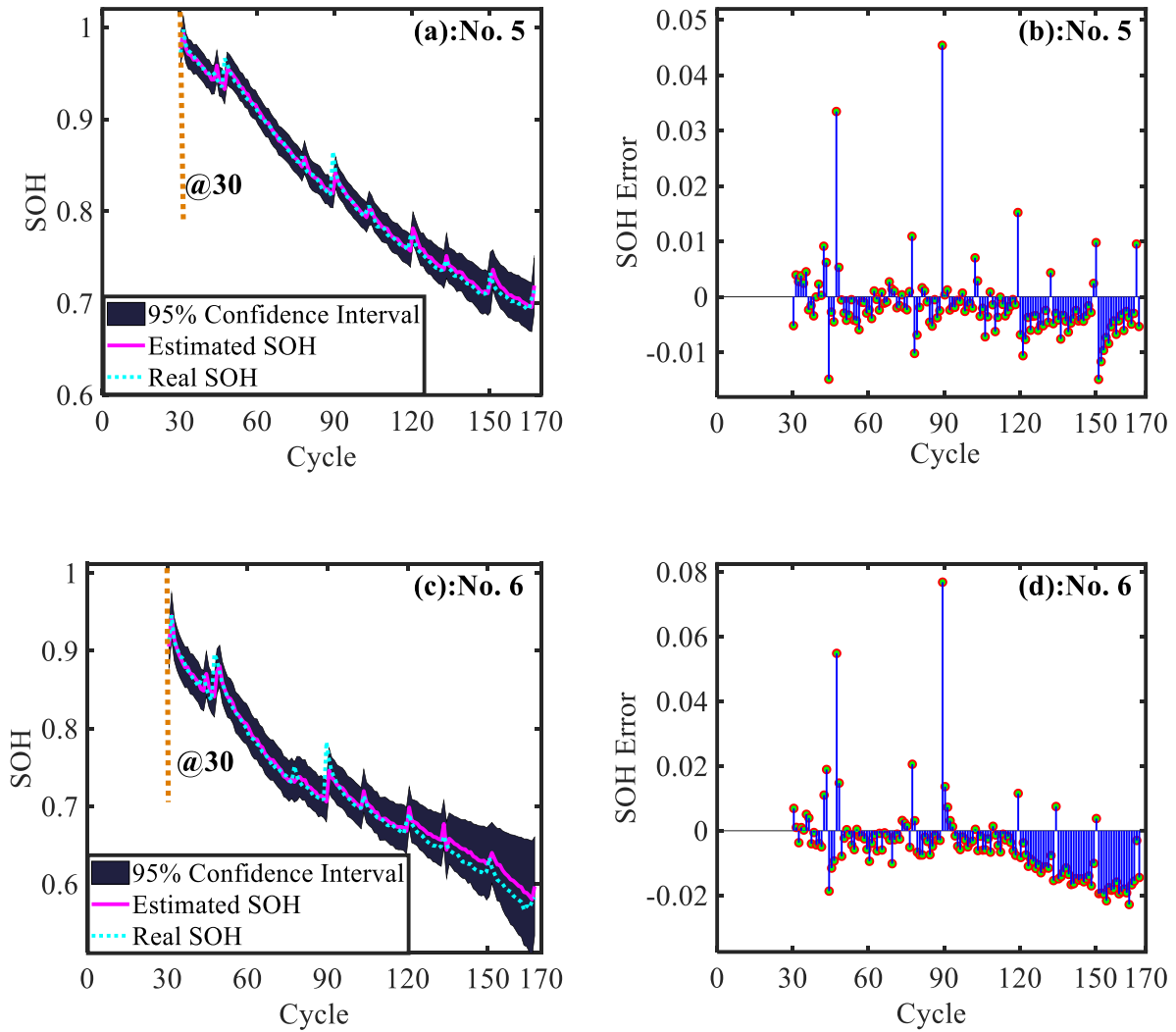

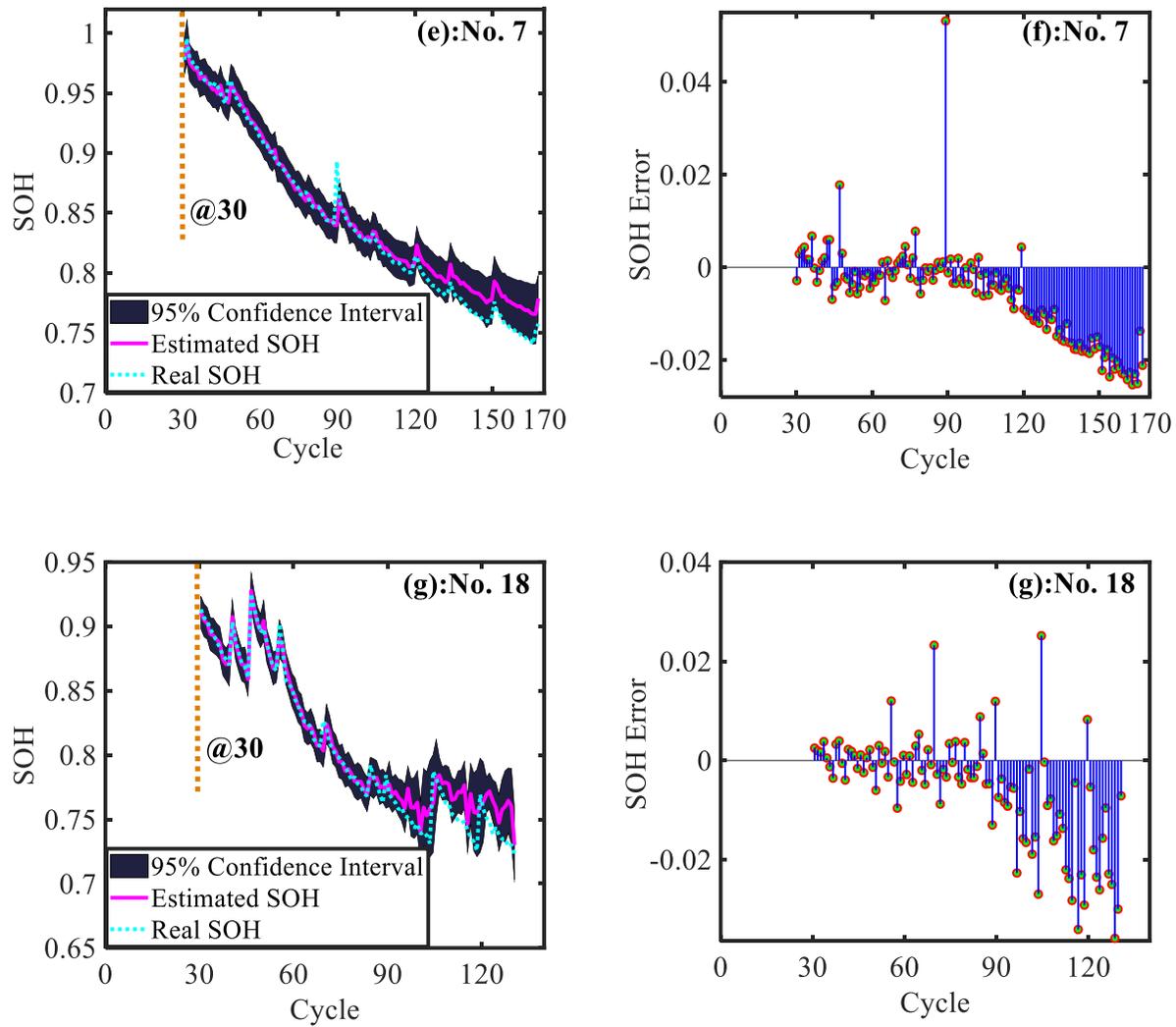

Fig. 7. The robustness analysis of battery SOH estimation based on GPR model: (a) SOH estimation result for battery No. 5. (b) SOH estimation error for battery No. 5. (c) SOH estimation result for battery No. 6. (d) SOH estimation error for battery No. 6. (e) SOH estimation result for battery No. 7. (f) SOH estimation error for battery No. 7. (g) SOH estimation result for battery No. 18. (h) SOH estimation error for battery No. 18.

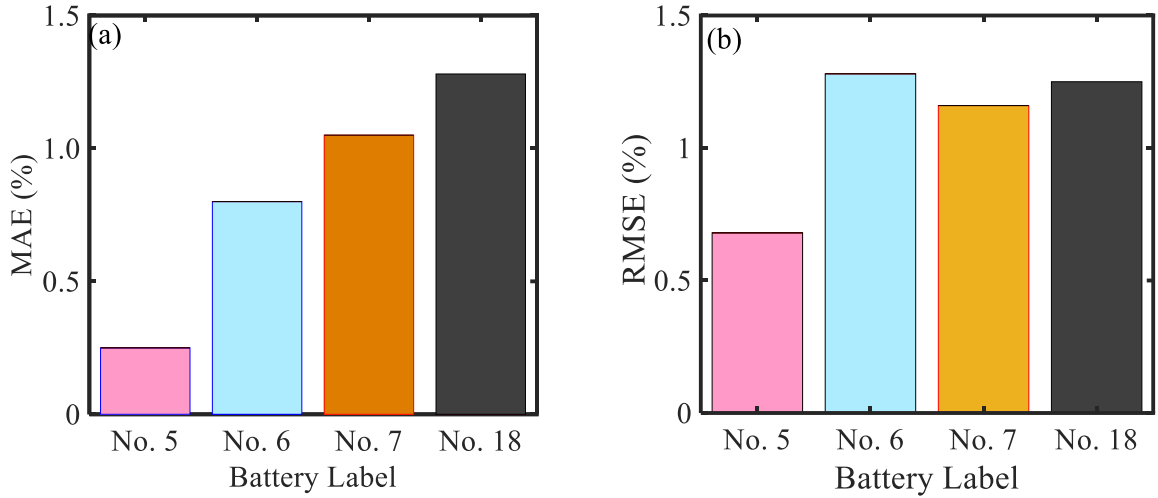

Fig. 8. The error analysis of battery SOH estimation under different initial conditions: (a) The MAE for the tested batteries. (b) The RMSE values for the tested batteries.

## V. CONCLUSION

This paper presents a Bayesian nonparametric approach for lithium battery SOH estimation based on the health indexes of partial charging IC curves. The main contributions are summarized as follows: (1) the IC curves are used to analyze the battery health levels during the charging process; (2) Compared the filtering results of the advanced Gaussian filter method with the moving average methods, the Gaussian filter method then is applied to smooth the IC curves; (3) The battery health indexes are extracted from the partial IC curves and regarded as the input data for the battery degradation model. (4) The GPR is employed to establish battery degradation model and demonstrate the model uncertainty. (5) The proposed model is verified by using normal battery degradation datasets and the different initial health conditions datasets. To sum up, the proposed GPR-based battery degradation model in this work has advantages of high accuracy and robustness. The drawback of this method maybe not suitable for other types of batteries due to the difference of charging IC curves. Thus the range selection procedure of the complete IC curves needs to change. The effectiveness of the proposed method was experimentally verified with a maximum estimation error of 3%. In the future, the current model should be tested in other different aging conditions for the sake of providing a more accurate, time-saving and universal battery degradation model should be studied.